\documentclass[10pt]{article}
\usepackage{ecmatex,aaspp4}
\usepackage{lscape}

\newcommand     {\HI}           {H\,{\small I}$\;$}
\begin{document}
\title{Supernova Remnant in a Stratified Medium:
Explicit, Analytical Approximations for Adiabatic Expansion
and Radiative Cooling}
\author{Witold Maciejewski}
\affil{Department of Astronomy, University of Wisconsin,
475 N. Charter St., Madison, WI 53706, \\
and Max-Planck-Institut f\"ur Astronomie, Heidelberg, Germany; 
witold@mpia-hd.mpg.de}
\author{Donald P. Cox}
\affil{Department of Physics, University of Wisconsin,
1150 University Ave., Madison, WI 53706; cox@wisp.physics.wisc.edu}
\begin{abstract}
We propose simple, explicit, analytical approximations for 
the kinematics of an 
adiabatic blast wave propagating in an exponentially 
stratified ambient medium, and for the onset of radiative
cooling, which ends the adiabatic era. Our method, based on
the Kompaneets implicit solution and the Kahn approximation
for the radiative cooling coefficient, gives straightforward 
estimates for the size, expansion velocity, and progression of
cooling times over the surface, when applied to supernova 
remnants (SNRs). The remnant shape is
remarkably close to spherical for moderate density 
gradients, but even a small gradient in ambient 
density causes the cooling time to vary substantially over 
the remnant's surface, so that for a considerable period there 
will be a cold dense expanding shell covering only 
a part of the remnant.
Our approximation provides an effective tool for identifying 
the approximate parameters when planning 2-dimensional 
numerical models of SNRs, the example of W44 being given in a 
subsequent paper.
\end{abstract}
\section{Introduction}
A good model of a supernova remnant's (SNR's) blast wave propagating 
through the ambient medium
allows us to derive, from the observed quantities, important 
properties of the medium, as well as the explosion energy.
The Sedov solution (1959) for adiabatic expansion into an isotropic 
medium usually serves as the prototype, and further, more realistic
models also assume isotropy (see e.g. Jun, Jones \& Norman 1996).
Nevertheless, in most cases of astrophysical interest (stellar
winds, Garcia-Segura \& Mac Low 1995; SNRs in vicinity of molecular 
clouds, Dohm-Palmer \& Jones 1996; galactic supershells, Maciejewski
et al. 1996; and starburst galaxies, Jogee, Kenney \& Smith 1998), 
the ambient medium is not isotropic. At the next level of 
approximation,
Kompaneets (1960) found an implicit solution for a blast wave 
propagating in an exponentially stratified ambient medium,
under the assumption that the post-shock pressure is uniform over the
surface of the remnant. An explicit form can be obtained
when one additional assumption is made, that
the shape of the shock wave can be approximated by a prolate 
ellipsoid. After an alternative presentation of the Kompaneets 
solution in \S2, we derive explicit analytical formulae for the 
size, velocity and the onset of cooling in \S3. A prolate ellipsoid
is a good approximation for all evolutionary
times prior to blowout (Kompaneets 1960), but we show in \S4 that
the Kompaneets approximation of uniform post-shock pressure loses
validity much earlier. In \S5 we show how our 
explicit approximation can be used in exploring the parameter space 
to identify values of interest in making 2-dimensional 
numerical models of SNRs.
\section{A simple method to derive basic Kompaneets conclusions}
Before we introduce our assumption, let us show a simple way of
re-deriving the basics of the original Kompaneets solution, using one 
of his conclusions, namely that in the direction perpendicular to 
the density gradient, the remnant is widest halfway between the 
high and low ambient density extremes of the shock, a location we
shall refer to as the equator. 
With the resulting formulae for the remnant size parallel and
perpendicular to the gradient, our approximation for the shock shape
immediately provides the volume and pressure needed to learn the time
evolution.

We consider the shock expansion in cylindrical coordinates $(z,r)$:
the explosion occurs at $(z,r) = (0,0)$ (Fig.1). The ambient 
density $\rho_0$ is an exponential function of $z$:
\begin{equation}
\label {e:A0}
\rho_0 = \rho_* e^{-z/h},
\end{equation}
where $\rho_*$ is the density at the explosion site 
and $h$ is the stratification scaleheight. Denoting $z_H$ and $z_L$ 
as the location of 
the dense and the tenuous end of the remnant, respectively, we
have $z_L-z_H = 2a$, where $2a$ is the extent of the remnant in the
direction along the density gradient. For a strong, non-radiative
shock in a $\gamma=5/3$ gas, the post-shock pressure $P_S$, pre-shock 
density $\rho_0$ and shock velocity $v$ are related by
\begin{equation}
\rho_0 v^2 = \frac{4}{3} P_S.
\label {e:A1}
\end {equation}
Assuming after Kompaneets that the post-shock pressure $P_S$
is constant over the surface of the remnant, 
the shock speed at a particular time can be written as a function 
of $z$ :
\begin{equation}
\label{e:A2}
v(z,t) = v_*(t) e^{z/2h}.
\end{equation}
\begin{figure}[t]
\vspace{-.5cm}
\plottwo{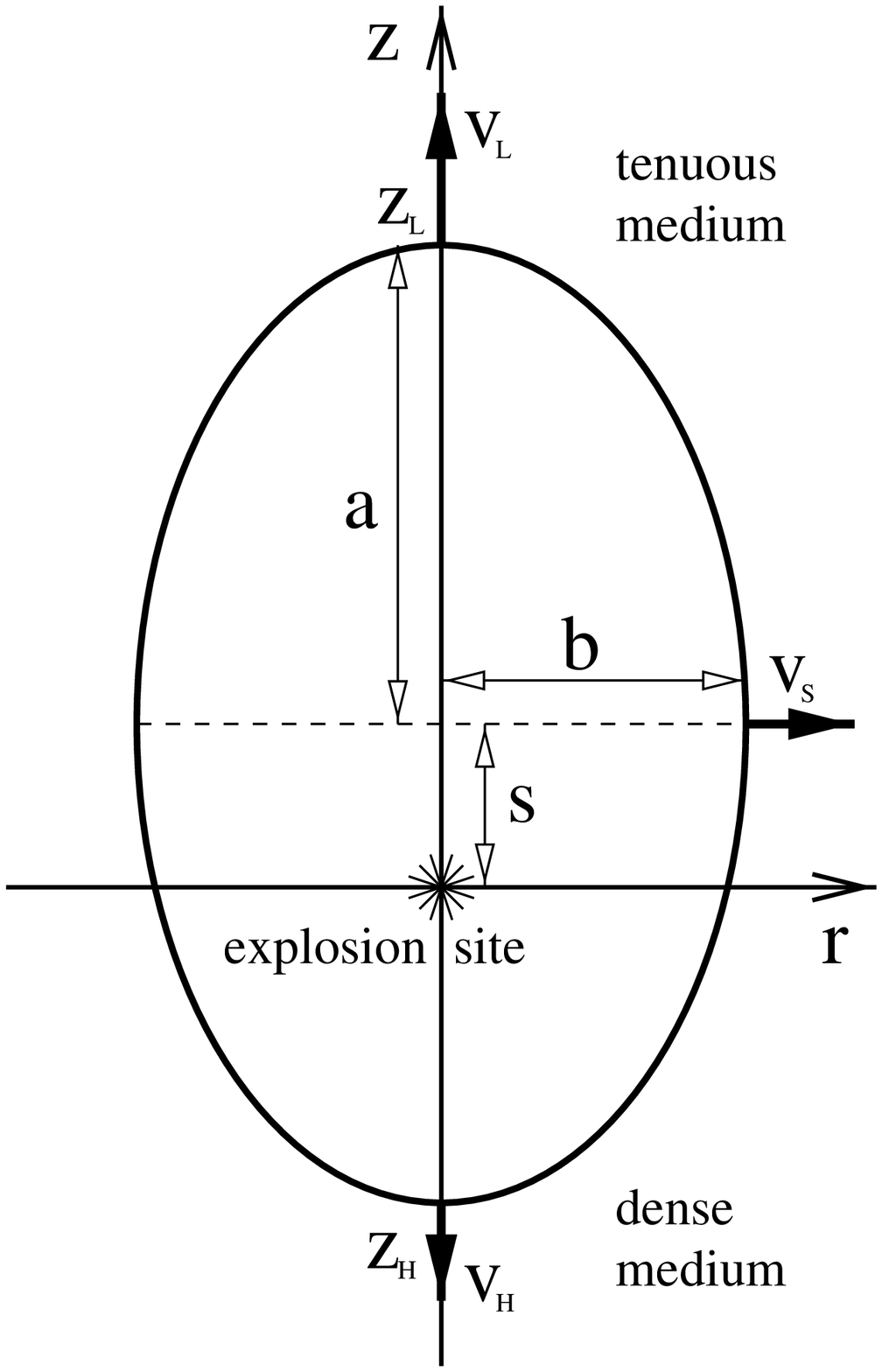}{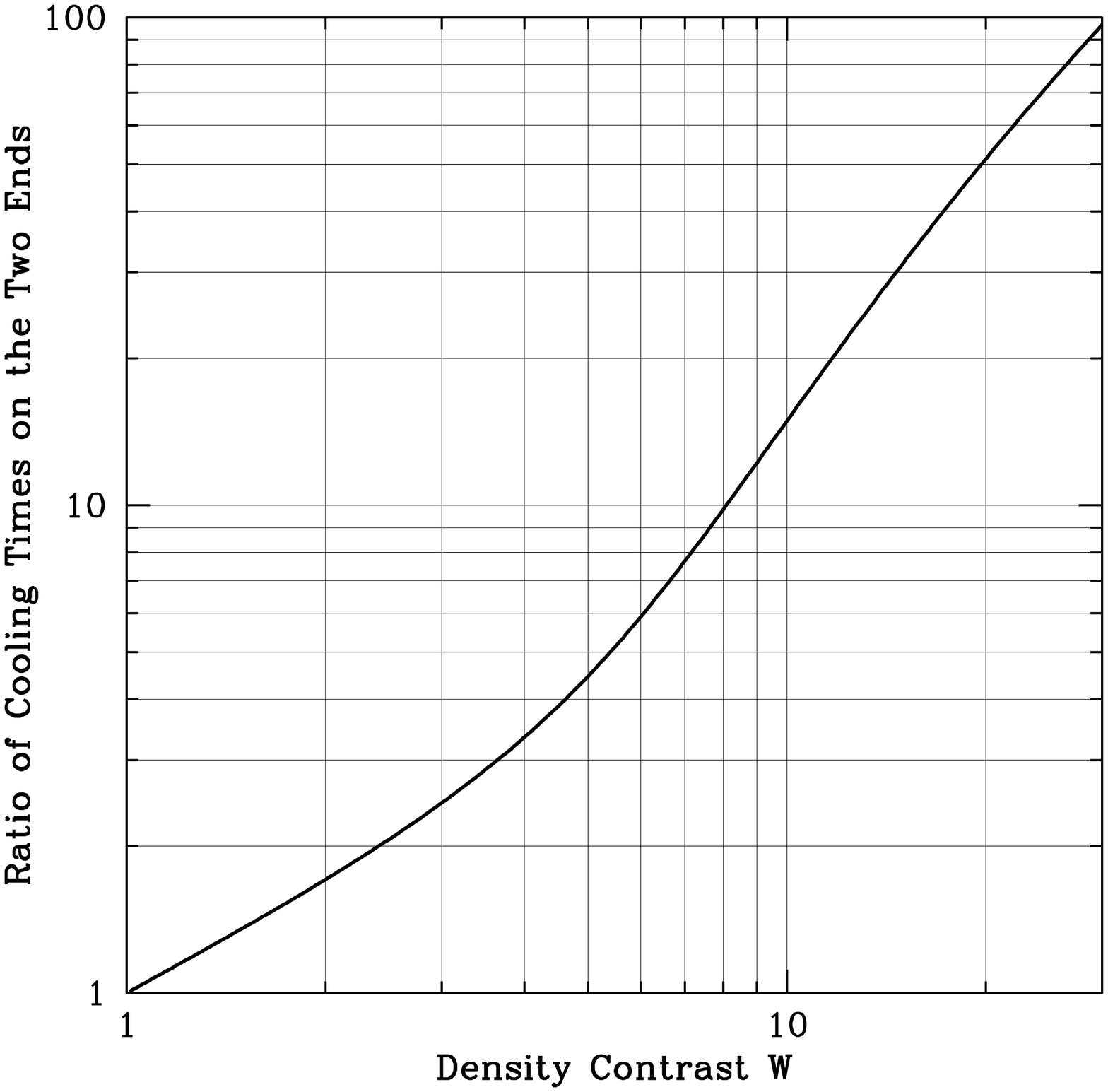}
\caption{(left) Representation of SNR characteristics used in the text. 
The ellipsoid is axially symmetric about the $z$ axis. Fig. 2.---
(right) The ratio of cooling times on the tenuous and dense ends of 
the remnant
expanding in a stratified medium, as a function of the ratio of extreme
ambient densities of an equivalent remnant at its cooling time. The
equivalent remnant is defined as a remnant with the same explosion 
energy, but expanding in a uniform density equal to the density at the 
explosion site. For the radius $r_{\rm cool}$ of the equivalent remnant
at its cooling time, the ratio of extreme ambient densities is defined 
as $W=\exp(2r_{\rm cool}/h)$, where $h$ is the density scaleheight of
the ambient medium.}
\end{figure}
Comparing the values of shock velocities at the ambient density
extremes and at the equator, we can eliminate the unknown $v_*(t)$ and
calculate the evolution of the remnant shape as a function of its
semi-major axis $a$.

The rate of expansion along the density gradient is an average of
shock velocities at two density extremes: $v_H$ and $v_L$
\begin{equation}
\label{e:A3}
\dot{a} = \frac{v_H+v_L}{2} = v_*(t) \frac{e^{z_H/2h}+e^{z_L/2h}}{2} ,
\end{equation}
where $v_H$ and $v_L$ are velocity magnitudes, $z_H$ is negative,
and $v_H=-\dot{z}_H$.
The center of the remnant is displaced from the
explosion site by $s = 0.5 (z_L + z_H)$. At $z=s$ (the equator), 
the remnant expands
only in the lateral direction and the expansion rate is
\begin{equation}
\label{e:A4}
v_S = v_*(t) e^{s/2h} .
\end{equation}
Although the equator moves along the $z$ axis (Fig.1) as the 
remnant expands, the rate of change of the remnant's semi-minor 
axis $b$ is exactly equal to $v_S$: $\dot{b}=v_S$. The evolution 
of the displacement $s$ can be written as
\begin{equation}
\label{e:A5}
\dot{s} = \frac{\dot{z}_L+\dot{z}_H}{2} = \frac{v_L-v_H}{2} = 
v_*(t) \frac{e^{z_L/2h}-e^{z_H/2h}}{2} .
\end{equation}
After substituting $z_H=s-a$ and $z_L=s+a$ and simple 
manipulations, we can eliminate the constant $v_*(t)$ and get
\begin{equation}
\label{e:A6}
\frac{da}{db} = \cosh \frac{a}{2h} \hspace{1cm} ; \hspace{1cm}  \frac{ds}{da} = \tanh \frac{a}{2h} .
\end{equation}
Integrating the equations above, we can readily find $b$ and $s$ as 
functions of $a$
\begin{equation}
\label{e:A7}
\tan \frac{b}{2h} = \sinh \frac{a}{2h} \hspace{1cm} ; \hspace{1cm} \exp \frac{s}{2h} = \cosh \frac{a}{2h} .
\end{equation}
These results are independent of the assumed shape of the blast wave
and the same as found by Kompaneets (1960), after his auxiliary 
variable $x$
is recognized as $\tanh(a/2h)$. In particular, the equations above
allow us to estimate the flattening of the remnant $b/a$ and the
relative displacement of the center $s/a$. It appears that even
remnants extending several scale-heights retain a nearly spherical
shape: for example, if $a=1.7h$, the ambient density contrast between
the ends of the shock is almost 30, but the remnant remains barely
flattened (size ratio $b/a$=0.9).
On the other hand, the ellipsoid's center is displaced from the 
explosion site by $s=0.38a$ in the low density direction. The same 
conclusions about sphericity and the center shift 
can be drawn from the analytical solution based on sectoral
approximation (Gnatyk 1988). Hydrodynamical models constructed by
Dohm-Palmer \& Jones (1996) arrive to these conclusions as well.
One can also notice
that $\exp(z_H/2h) = \exp((s-a)/2h) = 0.5 \left(1 + \exp(-a/h)
\right)$: in the blowout case, when $a \rightarrow +\infty$, it is
0.5. This is Kompaneets' famous result, that as blowout occurs and
$z_L$ becomes large, $z_H$ goes to a constant $-2h\log2$. The lateral
size at blowout also agrees with Kompaneets' prediction: for
$a \rightarrow +\infty$, we get $b/2h \rightarrow \pi/2$.
\section{Our approximations}
The implicit solution by Kompaneets shows that the shock has the 
shape of a somewhat boxy ellipsoid. In fact, its shape is virtually 
indistinguishable from a true ellipsoid within limits of accuracy 
set by Kompaneets assumption of 
uniform post-shock pressure. We use this fact in order to introduce 
time as an independent variable instead of $a$, when we assume that the
post-shock pressure $P_S$ has the same volume relationship as in the 
1D Sedov solution (see e.g. Bisnovatyi-Kogan \& Silich 1995)
\begin{equation}
\label{e:A8}
P_S = \frac{4 \pi \zeta^5}{25} \frac{E}{V},
\end{equation}
where $\zeta^5=2.025$, $E$ is the explosion energy and
 $V = \frac{4}{3} \pi a b^2$ is the volume of the remnant approximated 
by an ellipsoid. Combining equations (\ref{e:A0}), (\ref{e:A1}) 
and  (\ref{e:A8}),
and identifying $z=s$ along the minor axis of the ellipsoid we have
\begin{equation}
\label{e:A9}
\rho_* e^{-s/h} \; \dot{b}^2 = \frac{4\zeta^5}{25}\frac{E}{a b^2} .
\end{equation}
After expressing $s$ and $b$ as functions of $a$, we get
\begin{equation}
\label{e:A10}
\frac{da}{dt} = \frac{\zeta^{5/2}}{5h} \sqrt{\frac{E}{\rho_*}} \frac{\cosh^2(a/2h)}{\sqrt{a} \; \arctan \sinh (a/2h)} ,
\end{equation}
which, after integrating gives
\begin{equation}
\label{e:A11}
t(a) \ = \ \zeta^{-5/2} \left(\frac{E}{\rho_*}\right)^{-1/2} (2h)^{5/2} \ I(a/2h) \ \equiv \ t_{2h} \ I(a/2h) ,
\end{equation}
where $t_{2h}$ is the nominal time at which a Sedov remnant in
a homogeneous medium would reach a radius of $2h$, and the integral 
$I(x)$ is defined in Table 1 for numerical evaluation. The numerical
form of $t_{2h}$ is 
\begin{equation}
\label{e:A11a}
t_{2h} \ = \ 17.32 \ {\rm yr} \; (2h_{\rm pc})^{5/2} \ n_*^{1/2} / E_{51}^{1/2}
\end{equation}
where $h_{\rm pc}$ is $h$ in parsecs, $E_{51}$ is $E$ in units of 
10$^{51}$ergs, and $n_*$ is the nuclear number density of the ambient
medium in cm$^{-3}$.
In this notation, the mass density is $\rho=mn$, where the average 
mass per nucleus is $m=(1.4/1.1)m_H$, $m_H$ being the mass of hydrogen.

The shock velocity at any $z$ can be easily calculated by combining
equations (\ref{e:A1}) and (\ref{e:A8}) and substituting the local 
ambient density for $\rho_0$. 
In the particular case of top, bottom and equator expansion 
velocities, one can use equation (\ref{e:A3}), which after combining
with equation (\ref{e:A2}) gives 
$\dot{a} = v_H \exp(a/2h) \cosh(a/2h) = v_L \exp(-a/2h) \cosh(a/2h)$.
After substituting $\dot{a}$ from equation (\ref{e:A10}), one gets 
explicit formulae for shock velocities at two density extremes:
$v_H$ and $v_L$. Obviously, the shock expansion velocity on the
equator is the geometrical average of $v_H$ and $v_L$
--- these results 
are given in Table 1. The ratio of expansion velocities at the two
ends is $\exp(a/h)$.

In SNR evolution in a uniform medium, the adiabatic era is brought
to close by the onset of significant radiative cooling, at remnant
radius $r_{\rm cool}(E,\rho_*)$, closely 
followed by formation of a dense shell. When a remnant is evolving 
in a density gradient of scaleheight $h$, cooling occurs first on 
the dense end, and then spreads over the remnant surface. The 
evolution has two characteristic scales: the kinematic $h$ and
the radiative $r_{\rm cool}$, with a history of shell formation
that is dependent on their ratio.

In order to estimate cooling times,
we used Kahn's cooling law approximation (1976), in which the 
cooling coefficient takes the form $L = \alpha T^{-1/2}$. It is
a reasonable approximation to the actual cooling coefficient for
the temperature range of cooling SNRs (Smith \etal 1996), and it has 
the special property that in the absence of thermal conduction, the 
time for a parcel of hot gas to cool from an initial state is 
independent of its history and equals
\begin{equation}
\label{e:Deltat}
\Delta t_{\rm cool} = \frac{P_S}{Ln^2}.
\end{equation}
For a newly shocked parcel of gas, the post-shock values of pressure
$P_S$, temperature $T$, and density $n$ are known, so we can find
$\Delta t_{\rm cool}$. This incremental cooling time $\Delta t_{\rm 
cool}$ is added to the time at which a parcel is shocked, to find the
actual cooling time of that parcel, and then the sum minimized over all
parcels to find the earliest parcel to cool. For a homogeneous
medium, $\Delta t_{\rm cool}$ is proportional to $P_S^{3/2}$ and
therefore to $r^{-9/2}$ or $t^{-9/5}$. Numerically, the minimum of 
$t+Ct^{-9/5}$ occurs at $t_1=[9C/5]^{5/14}$ such that the first gas to 
cool was initially a distance 
\begin{equation}
\label{e:r1d}
r_1 \equiv r(t_1) \ = \ 19.62 \ {\rm pc} \; E_{51}^{2/7} / n_*^{3/7} 
\end{equation}
from the explosion site, and was shocked at time $t_1$
(see Cox \& Anderson 1982, Cox 1986). 
The expression for $C$ can be obtained by substituting equation 
(\ref{e:A1}) to (\ref{e:Deltat}), and using formulae for the Sedov radius
and velocity from Table 1. One can get then 
$ C = \frac{1}{32} \; \left( \zeta \sqrt{3}/5 \right) ^3  \; 
\left[ m^{9/10}/(\alpha \sqrt{\chi k}) \right]
 \; \left( E^{0.6}/n_0^{1.6} \right)$.
At the minimum, 
$\Delta t_{\rm cool}=\frac{5}{9}t_1$, for other radii,
$\Delta t_{\rm cool}=\frac{5}{9}t_1(r_1/r)^{9/2}$. 
First parcel's cooling is 
complete at $t_{\rm cool} = \frac{14}{9}t_1$, when the remnant radius
is $r_{\rm cool} = (\frac{14}{9})^{2/5} r_1$. We define 
$t_{\rm cool}$ as the cooling time of the 
remnant.

In our approximation for the remnant in a density gradient, a 
combination of equations (\ref{e:A0}), (\ref{e:A8}) and 
(\ref{e:Deltat}) gives
\begin{equation}
\label{e:A12}
\Delta t_{\rm cool}(a,z) = \frac{5}{9} \ t_1(E,n_*) \left(\frac{r_1^3}{ab^2}\right)^{3/2} \left(\frac{n_*}{n_0}\right)^{5/2} = \ 1.066\times10^{10} {\rm yr} \ \frac{E_{51}^{3/2}}{(2h_{\rm pc})^{9/2} n_*^{5/2}} \ \frac{\exp(5z/2h)}{(a/2h)^{3/2}(b/2h)^3} ,
\end{equation}
where we substituted 
$\alpha=1.3\times10^{-19}$ K$^{1/2}$cm$^3$erg s$^{-1}$ in Kahn's 
formula for cooling. Noting that $t_{2h} = t_1(2h/r_1)^{5/2}$ yields
\begin{equation}
\label{e:A12a}
\Delta t_{\rm cool}(a,z) = \frac{5}{9} \ t_{2h} \ \left(\frac{r_1}{2h}\right)^7 \ \frac{\exp(5z/2h)}{(a/2h)^{3/2}(b/2h)^3} .
\end{equation}
A measure of the likehood of finding a remnant with a cold
shell only on the dense end is provided by the ratio 
$t_{\rm cool}^-/t_{\rm cool}^+$ of cooling
times between the tenuous and dense ends.
These cooling times can be obtained by minimizing the 
value of the sum $\Delta t_{\rm cool} + t(a)$ of times given by 
equations (\ref{e:A12a}) and (\ref{e:A11}) over $x=a/2h$ 
\begin{equation}
\label{e:A12c}
t_{\rm cool}^{\pm} \ = \ t_{2h} \; {\rm min}_x \left[ I(x) \ + \ \frac{5}{9} (\frac{r_1}{2h})^7 \ g_{\pm}(x) \right] \ ,
\end{equation}
where the function $g_\pm(x)$ is given in Table 1, and the upper (lower)
sign corresponds to the dense (tenuous) end. The formula to derive 
values of $x$ at minimum, $x_c^\pm$, is given in Table 1. One can
see that $x_c^\pm$ is a function of only one variable: 
$r_1/2h = (14/9)^{-2/5} r_{\rm cool}/2h$.

The ratio of cooling times between the tenuous and 
dense ends of the remnant is numerically close to the density contrast 
between the two ends, $W=\exp(2r_{\rm cool}/h)$, at the nominal
cooling radius. Figure 2 displays the ratio of cooling times as a 
function of $W$. 
Note that for cooling time ratios less than 10 (essentially all for
which the parameter is interesting), it is equal to the density
contrast $W$ within
20\%. From Figure 2 we see that cooling takes more than twice as long 
on the tenuous end for $W>2.4$, or roughly speaking, for 
$r_{\rm cool}>h/2$, yielding the unsurprising result that when the 
density differential between the two ends is large at the nominal onset
of cooling, the evolution will contain a long period with a partial
shell. This result may
contribute to the explanation of the fact that we usually can see
only one side of the expanding \HI shells (Heiles 1979, Koo \& Heiles
1991).

For completeness, we now provide an approximate evaluation of the
time at which the first parcel on the equator cools, as a measure
of the proximity of a remnant to having half a complete shell.
This is somewhat more difficult, because the location of the equator
is shifting with $z$. Any parcel cooling at the equator was not
on the equator when shocked. Nevertheless, we are able to provide 
a plausible
approximation for the nearly spherical case. We use equation 
(\ref{e:A8}), which relates the post-shock pressure to the
explosion energy and volume, but approximate the volume by $V =
\frac{4}{3} \pi a^3$, and then substitute for $a$ the solution of the
Sedov 1D problem $a = (E/\rho_*)^{1/5} \zeta t^{2/5}$. The approximate
post-shock pressure is then
\begin{equation}
\label{e:A13}
P_S = \frac{3}{25} \zeta^2 (E^2 \rho_*^3)^{1/5} t^{-6/5} .
\end{equation}
From equations (\ref{e:Deltat}) and (\ref{e:A13}), the cooling time 
for a parcel on remnant's minor axis, which is hit by 
the shock at the time $t_s$, when the center of the remnant was at a 
distance $s$ from the explosion site, is
\begin{equation}
\label{e:A14}
\Delta t_{\rm cool}(s) \propto \frac{E^{3/5}}{\rho_*^{8/5}} \ t_s^{-9/5} \exp (5s/2h) .
\end{equation}
By assuming that $\Delta t_{\rm cool}(s) \approx \frac{5}{9} t_s$, which
is an exact equality for the uniform medium only, an approximate 
formula for $t_s$ can be derived.  The corresponding approximate 
expression for $t_{\rm cool}$ is
\begin{equation}
\label{e:A15}
t_{\rm cool} = \Delta t_{\rm cool}(s) + t_s = \frac{14}{9} t_s = 4.60\times10^4 {\rm yr}  \ \frac{E_{51}^{3/14}}{n_*^{4/7}} \ \exp ( \frac{25}{28} \frac{s}{h} ) .
\end{equation}
and approaches the cooling time for Sedov solution 
when $h\rightarrow +\infty$. The dependence on $s$ can be removed 
by expressing $s$ in terms of $a$ from equation (\ref{e:A7}) and 
substituting 
$a$ from the Sedov 1D solution. Thus we find the iterative solution 
for the cooling time on the minor axis
\begin{equation}
\label{e:A16}
t_{\rm cool} = 4.60\times10^4 {\rm yr}  \ \frac{E_{51}^{3/14}}{n_*^{4/7}} \ \cosh^{25/14} \ \left( 0.16 \left(\frac{E_{51}}{n_*} \right)^{1/5}  \frac{(\frac{9}{14}t_{\rm cool})^{2/5}}{h_{\rm pc}} \right).
\end{equation}
\section{Accuracy of the Kompaneets Approximation}
The assumptions of uniform post-shock pressure and an exponential 
atmosphere lead directly to an equation (\ref{e:A7}) for the 
displacement $s$ of the remnant center from the explosion site, 
as a function of the semimajor axis, $a$. In the 
hydrodynamical models performed for the particular case of SNR W44 
by Shelton \etal (1998, Paper II), we noticed that this equation 
appeared to overestimate the displacement of the center. 
That implies that the post-shock pressure at the high density end 
is higher than at the low density end.  In the following, we provide 
a rough justification for such a differential. Note that numerical 
studies of $s(a)$ may yield a reasonably useful test case for 
intercomparison of 2D hydrocodes.

The upturn in pressure at the edge of the Sedov remnant can be 
attributed to the effective gravity in the decelerating post-shock 
gas.  The post-shock mass velocity is $v_2 = \frac{3}{4} v$, so 
the outwardly directed effective gravity is 
roughly $-\frac{3}{4} \dot{v}$. 
Assuming that the post-shock gas is in hydrostatic equilibrium, we 
get $\frac{dP}{d\eta} \sim -\rho g \sim -3 \rho_0 \dot{v}$, given 
$\rho = 4\rho_0$ ($\eta$ is the coordinate locally perpendicular to
the shock front and directed outwards). The jump condition 
(\ref{e:A1}) however specifies 
the post-shock pressure, so the fractional gradient is
\begin{equation}
\label{e:A17}
\frac{1}{P} \ \frac{dP}{d\eta} \sim -4 \frac{\dot{v}}{v^2}.
\end{equation}
From the same jump condition we have
$\frac{\dot{P_S}}{P_S} = \frac{\dot{\rho_0}}{\rho_0} + 2 \frac{\dot{v}}{v}$.
For the exponential atmosphere, 
$\dot{\rho}_0 = \pm \rho_0 v/h$ at
the dense and tenuous ends of the remnant respectively. Thus 
\begin{equation}
\label{e:A18}
\frac{\dot{v}}{v} = \frac{1}{2} \left( \frac{\dot{P_S}}{P_S} - \frac{\dot{\rho}_0}{\rho_0} \right) = \frac{1}{2} \left( \frac{\dot{P_S}}{P_S} \mp \frac{v}{h} \right),
\end{equation}
or finally, at the ends of the remnant,
\begin{equation}
\label{e:A19}
\frac{1}{P} \frac{dP}{d\eta} \sim - 2 \left( \frac{\dot{P_S}}{P_Sv} \mp \frac{1}{h} \right).
\end{equation}
In this equation, the leading term on the right-hand side is 
the positive gradient term present in
the Sedov solution for a uniform medium. The second term is the 
correction factor we are looking for, positive on the high density 
end, negative on the low density end. It expresses the fact that the 
higher density end has a more rapid deceleration and therefore a 
steeper pressure gradient.

For the Sedov case, equation (\ref{e:A8}) gives 
$-\dot{P_S}/P_Sv = 3/R$. With the semimajor axis $a$ in place of 
$R$, we use equation (\ref{e:A19}) to estimate the relative 
gradients between the two ends at
\begin{equation}
\label{e:A20}
\frac{\left(\frac{1}{P} \frac{dP}{d\eta}\right)_{\rm Dense}} {\left(\frac{1}{P} \frac{dP}{d\eta}\right)_{\rm Tenuous}} \sim \frac{1 + \frac{a}{3h}}{1 - \frac{a}{3h}}.
\end{equation}
Upon integrating from the central pressure plateau up the slope 
along the major axis, the ratios of the post 
shock pressures should be similar to, but less 
extreme than, this ratio of gradients.  As a rough parameterization, 
we can write it as
\begin{equation}
\label{e:A21}
\frac{ P_{S,\rm Dense}}{ P_{S,\rm Tenuous}} \sim \frac{1 + \frac{a}{Qh}}{1 - \frac{a}{Qh}} \ ,
\end{equation}
where $Q$ is somewhat greater than 3. At the end of this section, we 
provide a way of estimating the value of $Q$ from numerical studies.

Now, we can see that the Kompaneets approximation of constant 
post-shock pressure holds only for $a \leq h$, i.e. at times well 
before blowout. Our approximation of the shock by an ellipsoid 
is indistinguishable from the Kompaneets solution at this stage
(the $\Delta \eta/\eta$ deviations smaller than 10$^{-3}$). 

The post-shock pressure can be expressed as a function of time only,
when $a$ in equation (\ref{e:A21}) is approximated by the 1D Sedov 
solution
\begin{equation}
P_S \propto (1 \pm (t/t_Q)^{2/5}) t^{-6/5}, 
\end{equation}
where the top sign
refers to the dense end, the bottom to the tenuous one, and
$(t/t_Q)^{2/5}$ has replaced $a/Qh$. Keeping in mind that we are 
looking for a first order correction, we used the fact that at 
early times $P_S \propto t^{-6/5}$. For a strong, non-radiative 
shock, 
$v \sim \sqrt{P_S/\rho_0} \ \propto \ e^{z/2h} \sqrt{1 \pm (t/t_Q)^{2/5}} \: t^{-3/5}$.
Taking into account that
$v_L = \frac{dz_L}{dt}$, but $v_H = - \frac{dz_H}{dt}$, we can write 
\begin{equation}
\label{e:A22}
-e^{-z_H/2h} dz_H \propto \sqrt{1 + (t/t_Q)^{2/5}} \: t^{-3/5} \: dt \hspace{5mm} ; \hspace{5mm} e^{-z_L/2h} dz_L \propto \sqrt{1 - (t/t_Q)^{2/5}} \: t^{-3/5} dt ,
\end{equation}
which after integrating gives
\begin{equation}
\label{e:A24}
\frac{ e^{-z_H/2h} - 1 }{ 1 - e^{-z_L/2h} } = \frac{\left(1 + (t/t_Q)^{2/5} \right)^{3/2} -1}{1-\left(1 - (t/t_Q)^{2/5} \right)^{3/2}} \simeq \frac{1+\frac{1}{4}(t/t_Q)^{2/5}}{1-\frac{1}{4}(t/t_Q)^{2/5}} \sim \frac{1+\frac{1}{4}\frac{a}{Qh}}{1-\frac{1}{4}\frac{a}{Qh}},
\end{equation}
where we first expanded the numerator and denominator to the two leading
terms, and then substituted $a$ from the Sedov solution. Note that for
uniform post-shock pressure, the ratio (\ref{e:A24}) is equal 1, and 
after substituting $z_H=s-a$, $z_L=s+a$, one gets the formula 
(\ref{e:A7}) for the center shift $s$.

Similarly, in the case of nonuniform post-shock pressure, we express 
$z_L$ and $z_H$ in terms of $a$ and $s$ and after
simple algebraic manipulations we get the final formula for 
the center shift correction
\begin{equation}
\label{e:A26}
\exp (s/2h) \ = \ \cosh (a/2h) - \frac{1}{4} \frac{a}{Qh} \sinh (a/2h) \ .
\end{equation}
The first order $s$ correction is about 4 times smaller than the $P_S$ 
correction in equation (\ref{e:A21}). The center shift correction derived 
above applies to models with no cooling; for numerical models, 
one could plot the quantity
\begin{equation}
\label{e:A27}
\frac{\cosh (a/2h) - \exp (s/2h)}{(a/2h) \sinh (a/2h)} = \frac{1}{2Q}
\end{equation}
to evaluate the free parameter $Q$ in our method.
\section{Applications to hydrodynamical models and conclusions}
Assuming that the shock wave propagating in an exponentially 
stratified medium takes the shape of an ellipsoid, we were able 
to find explicit expressions for its size and expansion velocity 
as functions of time, and for the cooling time at both ends and 
on the equator. As presented in Table 1, these expressions take 
forms similar to the Sedov (1959) solution for a
uniform ambient medium. Nevertheless, while the Sedov solution is
self-similar, there are no such solutions for nonuniform media.
The reason is that there are at least three independent dimensional
parameters (in our case $E$, $\rho_*$ and $h$). A successful
model of a SNR should adopt parameters that reproduce the observed
quantities: for example the size, expansion velocity and progress of 
shell formation. If a pulsar is seen, the age can be derived as well.
Our simple formulae provide quick ways to explore this 
3-dimensional parameter space.

A useful approach is to generate topographic (or contour) plots on
the $\rho_*$-$h$ plane for assumed values for $E$ and linear size of the 
remnant. The contoured quantities include the time required to 
reach the assumed size, the post-shock pressure, the cooling and 
shell formation time-scales for the dense, equatorial, and tenuous 
directions, and the shock velocities in those directions. Although 
the Kompaneets model, like the Sedov model, cannot be used very 
reliably to estimate quantities after shell formation has occurred, we
make the usual approximation that shortly after shell formation the 
shell velocity is about three-quarters of the shock velocity predicted for 
the non-radiative evolution. By this means of exploring the 
parameter space, a set of parameters fully reproducing the observables
for SNR W44 was found and is given in Paper II.

There are other asymmetric SNRs which can be interpreted in terms 
of expansion into a nonuniform medium, for example a partially 
open \HI shell in CTB 80 (Koo \etal 1990), and an incomplete shell
of CTA 1 (Pineault \etal 1993).
The semi-circular shape of CTB 109 is ascribed to interaction 
with a molecular cloud (Tatematsu \etal 1990). G 84.2 -0.8 (Feldt
\& Green 1993) shows a striking resemblance to W44 in radio
continuum, though observations of the \HI shell are not convincing.

In the catalogue of SNRs with \HI shell emission (Koo \&
Heiles 1991, Table 3) all sources show high-velocity \HI gas on
one side only: either receding or approaching. Heiles (1979), who 
catalogued \HI shells in our Galaxy, noticed 
that for most of them, he could see only the approaching or only the
receding hemisphere -- a fact that he called ``disturbing''. Putting 
aside possible observational biases, a partially formed shell appears
to be more a rule than an exception. 
We want to point out that explanation of such asymmetric SNRs or 
shells does not require any abrupt density change, such as encountering
the edge of a molecular cloud (Dohm-Palmer \& Jones 1996). The method
of estimation of the cooling time presented in \S3 shows that the
ratio of the cooling times on the dense and tenuous ends of the
remnant is of the order of the ratio of ambient
density extremes. We can see one side of the \HI shell only, because
at first there is only one side, and then by the time the other side
is formed, either the part expanding into denser medium has slowed 
down and its emission blends with that of local gas or the
tenuous end remains invisible due to the high column density contrast
between low and high density ends (Silich 1992).

Our explicit analytical approximation 
provides an efficient tool for a quick exploration of the space
of initial parameters for the SNR models. It can be used to select the
initial parameters of hydro runs,
which in turn can give us a detailed insight into the structure of the 
remnant, and can verify the above conclusions.

We would like to thank the referee, S.A. Silich, for valuable comments, and Linda Sparke for helpful suggestions on
the paper. DPC was supported in part by NASA Grant NAG5-3155 
to the University of Wisconsin.

\newpage

\begin{landscape}
\begin{center}
Table 1.\\
Approximate Formulae for a SNR Expanding in a Stratified Medium
\end{center}
\vspace{-.5cm}
\hspace{-3cm}
\begin{eqnarray*}
\hline\hline
{\rm SEDOV \; 1D} & & {\rm ELLIPSOIDAL \; 2D \; (EXPONENTIAL \; STRATIFICATION)} \\
\hline
& & \\
t \ = \ \zeta^{-5/2} \; \left(\frac{E}{\rho_*}\right)^{-1/2} r^{5/2} & \hspace{.1cm} {\bf size - time} \hspace{.1cm} & t \ = \ \zeta^{-5/2} \; \left(\frac{E}{\rho_*}\right)^{-1/2} (2h)^{5/2} \; \; I\left(\frac{a}{2h}\right) \\
 & & \\
t \ {\rm [yr]} = 17.32 \; \frac{r_{\rm pc}^{5/2} \; n_*^{1/2}}{E_{51}^{1/2}} & & t \ {\rm [yr]} = 17.32 \; \frac{(2h_{\rm pc})^{5/2} \; n_*^{1/2}}{E_{51}^{1/2}} \; I\left(\frac{a}{2h}\right) \\
& & {\rm where} \; I(x) \equiv \frac{5}{2} \int_{0}^{x} dy \frac{\sqrt{y}}{\cosh^2y}  \arctan \sinh y \ \simeq \ \frac{3x^{5/2}}{3+2x^{5/2}} \ {\rm for} \ x<3\\
\hline
 & & \\
v_s \ = \ \frac{2}{5} \; \zeta^{5/2} \; \left(\frac{E}{\rho_*}\right)^{1/2} r^{-3/2} & {\bf velocity - size} & v_s \ = \ \frac{2}{5} \; \zeta^{5/2} \; \left(\frac{E}{\rho_*}\right)^{1/2} (2h)^{-3/2} \; \; f_\pm\left(\frac{a}{2h}\right) \\
 & & \\
v_s \ {\rm [km/s]} = 22600 \, \left(\frac{E_{51}}{n_*}\right)^{1/2} r_{\rm pc}^{-3/2} & & v_s \ {\rm [km/s]} = 22600 \, \left(\frac{E_{51}}{n_*}\right)^{1/2} (2h_{\rm pc})^{-3/2} \; f_\pm\left(\frac{a}{2h}\right) \\
& & {\rm where} \; f_\pm(x) \equiv \frac{\exp(\mp x) \; \cosh x}{\sqrt{x} \; \arctan \sinh x} \hspace{1mm} ^{*}\\
\hline
 & & \\
t_{\rm cool} \ {\rm [yr]} = 4.6\times10^4 \; \frac{E_{51}^{3/14}}{n_{*}^{4/7}} & {\bf cooling \; time} & t_{\rm cool} \ {\rm [yr]} = 17.32 \; \frac{(2h_{\rm pc})^{5
/2} n_*^{1/2}}{E_{51}^{1/2}} \; \left[ I(x_c^\pm) \; + \; A \: g_\pm(x_c^\pm) \right] \\
 & & {\rm where} \; \; g_\pm(x) \equiv \frac{\cosh^5x \; \exp(\mp5x)}{x^{3/2} \, (\arctan \sinh x)^3} \hspace{1mm} ^{*} , \; A \; = \; 6.2\times10^8 \; \frac{E_{51}^2}{n_{*}^3 (2h_{\rm pc})^7}\\
 & & {\rm and} \; x_c^\pm \; {\rm is \; set \; by} \; A = \frac{5}{2} \frac{x^2 \, \exp(\pm 5x) \, (\arctan \sinh x)^4}{\cosh^6x \, \left(\pm 5 {\rm e}^{\mp x} + \frac{3 \cosh x}{2 x} + \frac{3}{\arctan \sinh x}\right)}  \hspace{1mm} ^{*}\\ \hline
\end{eqnarray*}

\noindent Units: $h_{\rm pc}$ [pc], $r_{\rm pc}$ [pc], $n_*$ [cm$^{-3}$], $E_{51}$ [10$^{51}$erg]

\noindent$^*$ the top signs refer to expansion into the dense medium, the bottom into the tenuous medium; for the shock expansion velocity in the equatorial direction, the exponential factor should be dropped from $f_\pm(x)$

\end{landscape}

\end{document}